\documentclass[acmlarge]{acmart}

\AtBeginDocument{%
  }

\setcopyright{acmcopyright}
\copyrightyear{2018}
\acmYear{2018}
\acmDOI{XXXXXXX.XXXXXXX}

\acmJournal{POMACS}
\acmVolume{37}
\acmNumber{4}
\acmArticle{111}
\acmMonth{8}

\usepackage[T1]{fontenc}
\usepackage{color}
\usepackage{caption}
\usepackage{graphicx}
\usepackage{subcaption}
\usepackage{hyphenat}
\usepackage{comment}
\usepackage{fancyvrb}
\usepackage{url}
\usepackage[ruled,vlined]{algorithm2e}
\usepackage{algorithmic}
\usepackage{listings}
\usepackage{multirow}
\usepackage{tikz}
\usepackage{marvosym}
\usepackage{epstopdf}
\usepackage{tcolorbox}
\usepackage[mode=buildnew]{standalone}
\usepackage{wrapfig}
\usepackage{fvextra}
\usepackage{soul}
\usepackage[normalem]{ulem}

\definecolor{light-gray}{gray}{0.80}
\lstdefinestyle{codestyle}{
  basicstyle=\fontsize{7}{7}\selectfont\ttfamily,
  keywordstyle=\color{magenta},
  breaklines=true,                 
  captionpos=b,
  numbers=left,                    
  numbersep=5pt,
  frame = single,
  framexleftmargin=15pt,
  xleftmargin= 22pt,
  xrightmargin=10pt
}
\newcommand{\fb}{\textit{FuSeBMC}}
\newcommand{\tracer}{\textit{Tracer}}

\newcommand\ie{\textit{i.e.\ }}

\newcommand{\ahmedhappy}[1]{{#1}}

\begin{document}

\title[\textit{FuSeBMC v4}: Improving code coverage with smart seeds via {BMC,} fuzzing and static analysis]
      {\textit{FuSeBMC v4}: Improving code coverage with smart seeds via {BMC,} fuzzing and static analysis}




\author{Kaled M. Alshmrany}
\email{kaled.alshmrany@postgrad.manchester.ac.uk}
\orcid{1234-5678-9012}
\affiliation{%
  \institution{The University of Manchester}
  \city{Manchester}
  \country{UK}
  \and
  \institution{Institute of Public Administration}
  \city{Jeddah}
  \country{Saudi Arabia}
}

\author{Mohannad Aldughaim}
\affiliation{%
  \institution{The University of Manchester}
  \city{Manchester}
  \country{UK}
  \and
  \institution{King Saud University}
  \city{Riyadh}
  \country{Saudi Arabia}
  }
  \author{Ahmed Bhayat}
\affiliation{%
  \institution{The University of Manchester}
  \city{Manchester}
  \country{UK}
  }
  \author{Lucas C. Cordeiro}
\affiliation{%
  \institution{The University of Manchester}
  \city{Manchester}
  \country{UK}
  \and
  \institution{Federal University of Amazonas}
  \city{Manaus}
  \country{Brazil}
  }

\renewcommand{\shortauthors}{K. M. Alshmrany et al.}

\begin{abstract}
Bounded model checking (BMC) and fuzzing techniques are among the most effective methods for detecting errors and security vulnerabilities in software. However, there \ahmedhappy{are still} shortcomings in detecting these errors due to the inability \ahmedhappy{of existent methods} to cover large areas in target code. We propose \textit{FuSeBMC} v4, a test generator that \ahmedhappy{synthesizes seeds with useful properties, that we refer to as \emph{smart seeds},} {to improve the performance of its hybrid fuzzer} \ahmedhappy{thereby} achieving high C program coverage. \textit{FuSeBMC} \ahmedhappy{works by first} analyzing and incrementally injecting goal labels into the given C program to guide BMC and Evolutionary Fuzzing engines. \textcolor{black}{After that, the engines are employed \ahmedhappy{for an initial period} to produce \ahmedhappy{the so--called} smart seeds.} \ahmedhappy{Finally, the engines are run again, with these smart seeds as starting seeds, in an attempt to achieve maximum code coverage / find bugs. During seed generation and normal running, the \emph{Tracer} subsystem aids coordination between the engines. This subsystem conducts additional coverage analysis and updates a shared memory with information on goals covered so far}.  Furthermore, the \emph{Tracer} evaluates \textcolor{black}{test-cases} dynamically to convert cases into seeds for subsequent test fuzzing. \ahmedhappy{Thus, the} BMC engine \ahmedhappy{can provide} the seed that \ahmedhappy{allows the} fuzzing engine to \ahmedhappy{bypass} complex mathematical guards (\ahmedhappy{e.g., input validation}). As a result, we received three awards for participation in the fourth international competition in software testing (Test-Comp 2022), outperforming all state-of-the-art tools in every category, including the coverage category.
\end{abstract}


\received{10 May 2022}
\received[revised]{15 December 2022}
\received[accepted]{29 January 2023}

\maketitle

\section{Introduction}

Fuzzing is one of the essential techniques for discovering software bugs and is used by major corporations such as Microsoft \cite{godefroid2012sage} and Google \cite{googleFuzzing}. Fuzzers construct inputs known as \emph{seeds} and then run the program under test (PUT) on these seeds. The goal is to discover a bug by causing the PUT to crash. A secondary but essential goal is to cover as many program branches as possible since a bug occurring on a branch cannot be discovered if the branch is not explored. Broadly, fuzzers can be categorized in three ways. Firstly, blackbox fuzzers do not analyze the target program when generating seeds. Secondly, whitebox fuzzers extensively analyze the target programs to guide the seed generation to explore particular branches. Lastly, greybox fuzzing uses limited program analysis and feedback from the program under test to guide the input generation. \textcolor{black}{There are also hybrid fuzzers combining a concolic executor with a coverage-guided fuzzing approach. A scheduling and synchronization mechanism synchronizes them in a coordination mode.}

The main disadvantage of blackbox fuzzers is that {due to the random manner in which they generate inputs}, they are often unable to explore program paths with complex guards. Whitebox fuzzers, on the other hand, are very good at using program information to circumvent guards but are often slow and resource-intensive to run. Greybox fuzzing techniques such as the American Fuzzing Loop~\cite{americanfuzzylop_2021} offer a sweet spot regarding effort per input. However, they still have some fundamental weaknesses; most importantly, the straightforward way they generate seeds can lead to the fuzzer becoming stuck in one part of the code and not exploring other branches. Hybrid fuzzing attempts to {circumvent this issue with more significant program-specific analysis.} One common technique is \emph{concolic fuzzing}, which involves using a theorem prover to solve path constraints and {thereby helps the fuzzer to explore deeper into the program}~\cite{10.1145/3182657,mi2020shfuzz,stephens2016driller}.

This paper presents \emph{FuSeBMC}, a state-of-the-art hybrid fuzzer incorporating various innovative features and techniques. This journal paper is based on several published conference papers \cite{alshmrany2020fusebmc,alshmrany2021fusebmc,alshmrany2022fusebmc}. { In practice, we concentrated on the enhancements made to \emph{FuSeBMC} between 2021 (when our TAP paper \cite{alshmrany2021fusebmc} was published) and 2022. \textcolor{black}{In this journal paper, we expand on these enhancements, such as using the Tracer subsystem, shared memory, and analyzing and ranking goals. In addition, we demonstrate the advancement achieved by carrying out a more thorough experimental evaluation}.} \ahmedhappy{To summarise,} we extend those papers by (i) discussing \textit{FuSeBMC} in greater detail (Section \ref{sec:ProposedMethod}) (ii) providing more examples, and (iii) providing a thorough and up-to-date experimental evaluation of the tool (Section \ref{sec:ExperimentalEvaluation}).

One of the primary features of \emph{FuSeBMC} is the linking of a greybox fuzzer with a \emph{bounded model checker}. A bounded model checker works by treating a program as a state transition system and then checking whether there {exists a path in this system of length less than some bound $k$} that violates the property to be verified \cite{Biere09,CordeiroFM12}. In this work, we use ESBMC, an efficient bounded model checker for C++ and C with support for checking many safety properties fully automatically~\cite{CordeiroFM12}. ESBMC works by translating the property to check and the bounded transition system into quantifier-free first-order logic. SMT-solvers are then run to find a model for $\neg P$ (the negated property) and $C$ (the translated transition system). {Finally, if a model is found, a counterexample is extracted, representing the set of assignments required to violate the property.}

Bounded model checkers such as ESBMC are now mature software, used industrially~\cite{gadelha2018esbmc} and capable of finding bugs in production software. We leverage this power of model checkers as a method for seed generation. During greybox fuzzing, if a particular branch has not been explored, ESBMC can provide a model (set of assignments to input variables) that will reach the branch. This model is then used as a seed for further greybox fuzzing. {We evaluate seeds based on two criteria—the depth of the seed's deepest goal and the number of goals covered specifically by the seed. Smart seeds are those that score high on these metrics. The technique is implemented in \emph{FuSeBMC} \cite{alshmrany2020fusebmc}, and it is available for download from GitHub\footnote{\url{https://github.com/kaled-alshmrany/FuSeBMC}}.}

An important \emph{FuSeBMC} subsystem discussed in this paper is the \emph{Tracer}, which coordinates the bounded model checker and the various fuzzing engines. The {Tracer} monitors the \textcolor{black}{test-cases} produced by the fuzzers. It selects those with the highest impact (as measured by a couple of metrics discussed in Section~\ref{sec:ProposedMethod}) to act as seeds for future fuzzing rounds. Further, as discussed above, ESBMC produces \textcolor{black}{test-cases} to cover particular branches. However, a \textcolor{black}{test-case} it produces may also cover branches other than the one targeted. To ascertain precisely which branches a \textcolor{black}{test-case} covers and thereby prevent ESBMC from running multiple times unnecessarily, the Tracer takes a \textcolor{black}{test-case} produced by ESBMC and runs the PUT on it, recording all goals covered.

Bounded model checking can be slow and resource-intensive. To mitigate against this, \emph{FuSeBMC} does not use an off-the-shelf fuzzer for its {grey} box fuzzing but instead uses a modified version of the popular American Fuzzy Lop tool. One of the features of this modified fuzzer is its ability to carry out lightweight static analysis of a program to recognize input verification. It analyzes the code for conditions on the input variables and ensures that seeds are only selected if they pass these conditions. This reduces the dependence on the computationally expensive bounded model checker for finding quality seeds. Another interesting feature of the modified fuzzer is that it analyses the PUT and identifies potentially infinite loops heuristically. It then bounds these loops in an attempt to speed up fuzzing. These bounds are incremented during the multiple fuzzing rounds.

Together, these features turn \emph{FuSeBMC} into a leading fuzzer. In the 2022 edition of the Test-Comp software testing competition, \emph{FuSeBMC} achieved first place in both the main categories, \emph{Cover-Error} and \emph{Cover-Branches}. In the \emph{Cover-Branches} category, it achieved first place in 9 of the 16 subcategories it participated in. In the \emph{Cover-Error} category, it achieved first place, or joint first place, in 8 out of the 14 subcategories that it participated in.\\

\noindent\textbf{Contributions.}
This journal paper explains the latest developments to the \emph{FuSeBMC} fuzzer. The work presented here substantially extends our previous published conference papers~\cite{alshmrany2020fusebmc,alshmrany2021fusebmc,alshmrany2022fusebmc}. \emph{FuSeBMC}’s main new features can be summarised as follows:
\begin{itemize}
    \item \ahmedhappy{The use of lightweight static analysis to recognize some forms of input validation on variables, thereby enabling fuzzing to produce more effective seeds and speed up the fuzzing process.}
    
    
    \item The prioritization of deeper goals with regards to finding \textcolor{black}{test-cases} as this can result in providing higher code coverage and generating fewer \textcolor{black}{test-cases}.
    
    \item The setting of a loop unwinding depth during seed generation and fuzzing. As loop unwinding leads to exponential path explosion, we restrict the unwinding depth of each loop to a small number, depending on an approximate estimate of the number of program paths.
    
\end{itemize} 

\ahmedhappy{We also extend our previous papers by:}

\begin{itemize}
    \item Explaining the working of the \fb{} tool in greater depth and clarity than previously.
    \item Providing a detailed analysis of our participation in the international competition on software testing (Test-Comp 2022), where our tool \emph{FuSeBMC} achieved three significant awards. \emph{FuSeBMC} earned first place in all the categories by the improvements described in this manuscript. We also thoroughly compare version 4 of the tool and the previous iteration, version 3, thereby demonstrating the effectiveness of our extensions.
\end{itemize}

\section{Preliminaries}
\label{sec:Background}

This section briefly introduces various fuzzing techniques, including general \textcolor{black}{greybox} fuzzing, white box fuzzing, and hybrid fuzzing. In addition, it explains the various code coverage metrics in a simplified manner. Lastly, it introduces the technique of bounded model checking {and its application to automated test generation}.

\subsection{Fuzzing}
\label{sec:Fuzzing}
Fuzzing is one of the most effective software testing techniques for finding security vulnerabilities in software systems. It automatically generates inputs, i.e., \textcolor{black}{test-cases}, passes them to a target program, and checks for abnormal behavior or code coverage. The generated input should be well-formed and accepted by the software system to be considered a valid \textcolor{black}{test-case}. The inputs’ structure can be, for example, {very long or completely blank strings, min/max values of integers (or only zero and negative values), unique values, or characters likely to trigger bugs.}

\begin{figure*}
    \centering
    \includegraphics[width=\textwidth]{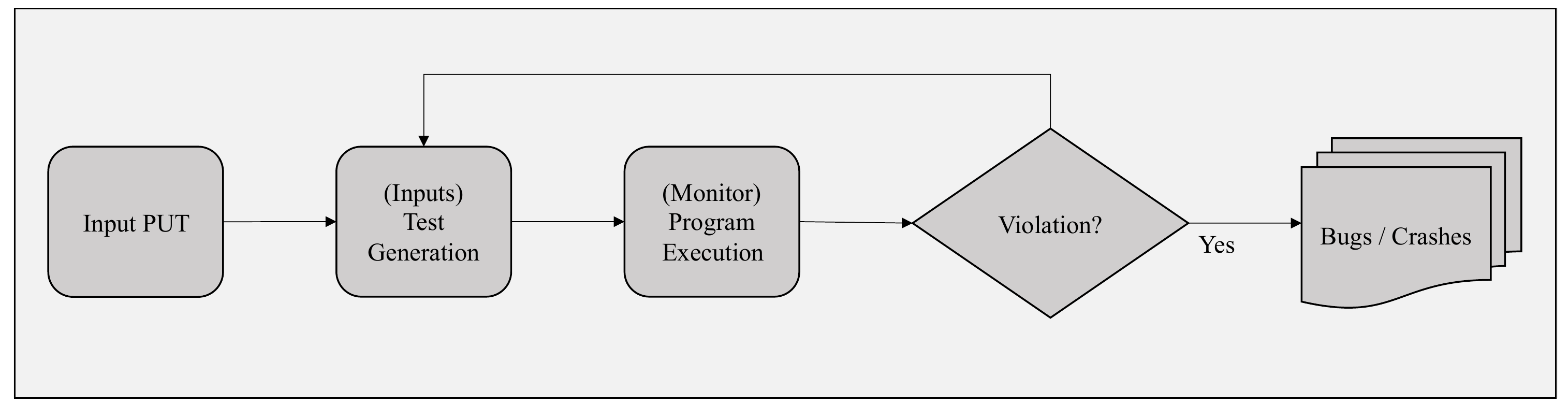}
    \caption{Traditional fuzzing {workflow}. The {fuzzing} process {consists of repetitive generation of inputs for the PUT, executing the PUT with these inputs and reporting any crashes as they occur.}}
    \label{figure:FuzzingOverview}
\end{figure*}

{Figure \ref{figure:FuzzingOverview} illustrates the traditional fuzzing approach} \cite{li2018fuzzing}. {It comprises two main phases: 1) input/test generation and 2) program execution and monitoring.} Generally, {fuzzing} starts with the target program and initial inputs, which can be any file format (e.g., images, text, videos, etc.), network communication data, and executable binaries. Generating malformed inputs is the main challenge for fuzzers. So, commonly, two kinds of generators are employed in state-of-the-art fuzzers, generation-based, which generates the inputs from scratch, or mutation-based, which modifies existing inputs. Inputs are fed to target programs after being generated in the previous step. \textcolor{black}{Then, the execution state is monitored by the fuzzer during the execution of the PUT to detect crashes or abnormal behaviors. When the fuzzer detects a violation, it stores the related \textcolor{black}{test-cases} for later usage and analysis.}

As the primary goal of fuzzing is to find more crashes with applicable inputs, the most intuitive performance metric is the rate of crashes per time. However, crashes rarely occur, so many existing fuzzers are designed to maximize code coverage. {They perform additional lightweight instrumentation to the PUT source code to enable coverage monitoring.} By maximizing the code coverage, the fuzzer will test more paths in the program, which increases the {likelihood} of causing crashes~\cite{kim2020maxafl}. Many undiscovered crashes occur in deep code paths, in parts of codes that are not executed frequently. Therefore, one of the critical roles of fuzzing is finding inputs that increase the code coverage. {The above briefly outlines the principles of coverage guided \textit{grey-box fuzzing}.}

{\textit{White-box fuzzing} aims to aid the fuzzing process by extracting some additional information about the PUT structure.} It combines fuzz testing with symbolic execution~\cite{godefroid2008automated}. {A white-box fuzzer} symbolically executes the PUT with initial concrete inputs tracking every conditional statement, producing constraints over those inputs. The collected constraints capture how the program uses its inputs. The fuzzer then systematically negates each constraint and solves them using a constraint solver, creating new inputs that exercise different program paths~\cite{godefroid2008grammar}. \textcolor{black}{For example, consider a symbolic variable $i$ with an initial test-case where $i$ is set to $0$. If this is followed by a branch condition such as ``if $i= 10$ then'', the symbolic execution process would generate a constraint $i \neq 10$ resulting from the execution.} A constraint solver can solve this formula and obtain a concrete value that would satisfy this new path by negating this constraint. This process can be repeated for the newly created inputs and can be used to maximize code coverage.

\subsection{{Code Coverage}}
\label{sec:CoverageCode}

One of the challenges in software testing is assigning a quality measure to \textcolor{black}{test-cases} such that higher-quality \textcolor{black}{test-cases} detect more bugs than low-quality \textcolor{black}{test-cases}  \cite{ivankovic2019code}. These quality measures are called \textit{test adequacy criteria}, and \textit{code coverage} is the commonly used test adequacy criterion. Generally, code coverage is a measure of identifying the parts of PUT that execute when running the program and the degree to which a \textcolor{black}{test-case} covers the PUT.
{Choosing an appropriate coverage criterion is paramount to successful test generation \cite{7272926,6823879}. One of the research directions in this area involves moving from utilizing standard coverage criteria towards designing generic specification languages (e.g., Hyperlabel Test Objectives Language \cite{7927998}) with higher expressive power for defining test objectives. However, one of the main challenges in applying such techniques is their unavailability in most off-the-shelf automated test generation tools.}

{
\textcolor{black}{Some of the most popular coverage criteria supported by most test generators include \textit{statement coverage}, \textit{branch coverage}, and \textit{function coverage}. Statement coverage determines how many program statements have been executed at least once. Branch coverage tracks how many conditional program branches (\textit{if} statements, \textit{switch} statements, loop statements) have been visited. Function coverage measures how many functions (out of the total number of functions in the PUT) have been invoked at least once during the program execution.}}

\subsection{Bounded Model Checking}

{Model checking \cite{ClGP99-Mc} is a general verification technique that aims to determine whether a mathematical abstraction (in the form of a finite state transition system) of the underlying system satisfies the given specification (a set of properties formalized using temporal logic). A model-checking algorithm explores all reachable states of the system and checks whether the given properties hold in every state. If one of the properties fails, a \textcolor{black}{counterexample} -- a sequence of system states leading to the failure -- is produced.}

{
Bounded model checking \cite{Biere09} solves a similar problem but for the bounded executions of the system to be verified. In other words, given a positive $k$, a bounded model checking algorithm tries finding a \textcolor{black}{counterexample} of maximum length $k$. In practice, if no such \textcolor{black}{counterexample} can be found, the value of $k$ can be increased until one of the given properties is violated or the verification problem becomes intractable.}

{In software verification \cite{CordeiroFM12,10.1007/978-3-540-24730-2_15}} bounded model checking {is typically accompanied by a symbolic execution of the given program up to the user-defined positive bound $k$. The obtained bounded symbolic traces are} then automatically translated into a first-order logic formula $C$. {It is used with} a formula $P$ representing a property that needs to be checked in the verified program {to formulate a satisfiability problem $C \wedge \neg P$}. A counterexample to $P$ has been discovered if this formula is satisfiable. Otherwise, it can be concluded that $P$ holds up to the given context-bound $k$ in the given program. {In general, such satisfiability problems are solved by SAT/SMT solvers \cite{10.1007/978-3-540-78800-3_24,10.1007/978-3-319-08867-9_49,DBLP:journals/jsat/NiemetzPB14}. Many verification tools employing bounded model checking \cite{gadelha2019esbmc,kroening2014cbmc} are capable of verifying functional correctness (i.e., program assertions) and code reachability, as well as automatically checking some implicit properties such as the absence of buffer overflows, dangling pointers, deadlocks, and data races. The exact set of such properties varies depending on the chosen verification tool.}

{In the context of automated test generation for software, bounded model checking is used for obtaining a sequence of concrete input values (as well as a concrete thread schedule for multi-threaded programs) that allows reaching a predefined location (program statement) within the program. The desired testing objectives determine the choice of such program locations.}

\subsection{Hybrid Fuzzing {Techniques}}
\label{sec:HybridFuzzing}

In recent years, hybrid fuzzing {techniques} have been widely researched and discussed in the {software} security field \cite{stephens2016driller,alshmrany2022fusebmc,9955513}. Since fuzz testing alone fails to explore the complete program state space, it is often combined with a complementary verification technique such as bounded model checking or \textcolor{black}{concolic execution~\cite{sutton2007fuzzing}}. \textcolor{black}{Hybrid fuzzing typically involves combining multiple fuzzing techniques, each with its strengths and weaknesses, to achieve better results. Some common fuzzing techniques that may be integrated into hybrid approaches include \textcolor{black}{Generational Fuzzing, Mutation-Based Fuzzing,} Symbolic Execution, Concolic Testing, and Grammar-Based Fuzzing.}


{The main motivation behind combining such complementary techniques is to leverage the strengths of concolic execution and bounded model checking in generating inputs satisfying complex branch conditions, which are challenging to derive for mutation-based fuzzing.} At the same time, fuzzing can quickly {explore} deep paths with simple checks that {can offset} the large resource consumption of concolic execution and {bounded model checking}.

\section{FuSeBMC v4 Framework}
\label{sec:ProposedMethod}

\textit{FuSeBMC} combines dynamic and static verification techniques for improving code coverage and bug discovery. It utilizes the Clang compiler~\cite{CLANG} front-end to perform various code transformations, ESBMC (Efficient SMT-based Bounded Model Checking)~\cite{gadelha2020esbmc,gadelha2019esbmc} as a BMC and symbolic execution engine, and a modified version of the American Fuzzy Lop (AFL) tool~\cite{bohme2017directed,americanfuzzylop_2021} as well as a custom selective fuzzer ~\cite{alshmrany2021fusebmc} as fuzzing engines.

\begin{figure}
    \centering
    \includegraphics[width=\textwidth]{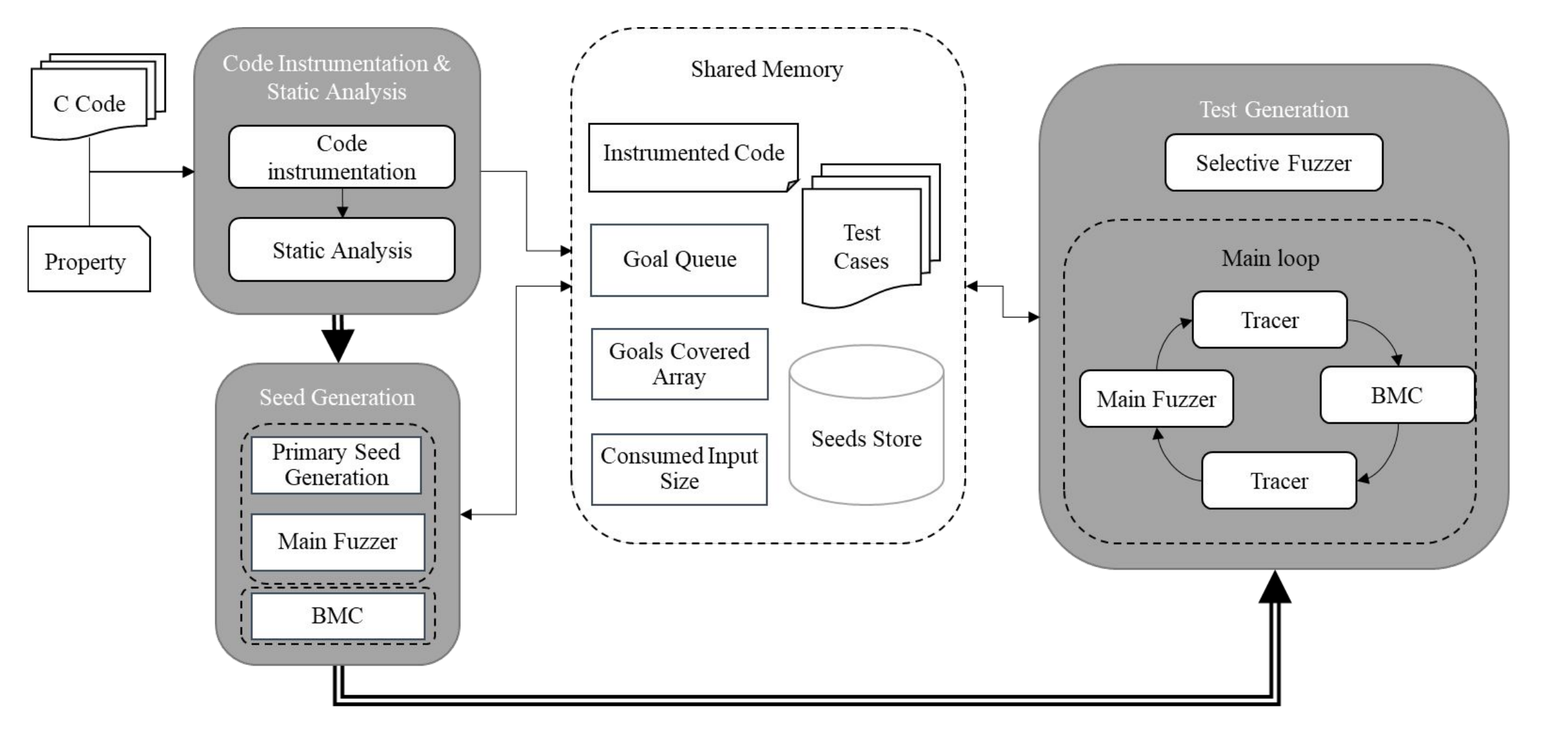}
    \caption{The Framework of \textit{FuSeBMC} v4. This figure illustrates the main components of \textit{FuSeBMC}. Our tool starts by instrumenting and analyzing the source code, then performs coverage analysis in two stages: seed generation and test generation.}
    \label{figure:FuSeBMCFramework}
\end{figure}

\textit{FuSeBMC} takes a C program as input and produces a set of \textcolor{black}{test-cases}, maximizing code coverage while also checking for various bugs. Users {can choose to check for several} types of bugs {that are supported by ESBMC} (such as array bounds violations, divisions by zero,  pointers safety, arithmetic overflows, memory leaks, and other user-defined properties). Figure~\ref{figure:FuSeBMCFramework} illustrates {the \textit{FuSeBMC} architecture and its workflow}, and Algorithm \ref{alg:FuSeBMC} presents the main stages of the \textit{FuSeBMC} {execution}.

\subsection{\bf Overview} 

{\em FuSeBMC} \textcolor{black}{begins by analyzing C code and then injecting goal labels into the given C program {(based on the code coverage criteria that we introduce in Section \ref{sec:CodeInstrumentation})} and ranking them according to {one of the} strategies {described in Section \ref{sec:GoalsRanking}} (i.e.,\ depending on the goal's origin or depth in the {PUT}).} From then on, {\em FuSeBMC}'s workflow can be divided into two main stages: {\em seed generation} (the preliminary stage) and {\em test generation} (the full coverage analysis stage). During {\em seed generation}, {\em FuSeBMC} applies the fuzzers and BMC to the instrumented code {once} for a short time to produce seeds that are used by the fuzzers at the {\em test generation} stage and \textcolor{black}{test-cases} that may provide coverage of some ``shallow'' goals. {The intuition behind this divide is to quickly generate some meaningful seeds for the fuzzer that could increase the chances of exploring the PUT past the entry point, which often contains restrictive input validators that are hard to negotiate for the fuzzers.} During {\em test generation}, the above engines are applied with a longer timeout while accompanied by another analysis engine called \tracer{}. It {helps} the execution of the fuzzers and the bounded model checker by recording which goal labels in the {PUT} have been covered by the \textcolor{black}{test-cases} produced by these engines. This is done to prevent the computationally expensive BMC engine from trying to reach an already covered goal. {\em FuSeBMC} {continues with the \textit{test generation} stage} until all goals are covered or a timeout is reached.

\begin{figure}
    \centering
    \begin{subfigure}{0.35\textwidth}
        \begin{lstlisting}[language=c,frame=none,xleftmargin=0ex]
#include <assert.h>
void reach_error() { }

bool check(int a, int b, int c, int x) {
 return (a*x*x + b*x + c == 0);
}

int main() {
 int a = __VERIFIER_nondet_int();
 int b = __VERIFIER_nondet_int();
 int c = __VERIFIER_nondet_int();
 if(b*b >= 4*a*c) {
  while(1) {
   int x = __VERIFIER_nondet_int();
   if(x <= 0 || x > 100) 
    reach_error();
   if(check(a, b, c, x)) 
    return 0;
  }
 }
 else
  reach_error();
 
return 0;
}
        \end{lstlisting}
        \caption{}
        \label{figure:original_code}
    \end{subfigure} %
    \begin{subfigure}{0.35\textwidth}
        \begin{lstlisting}[language=c,frame=none,xleftmargin=0ex]
#include <assert.h>
void reach_error() {
 GOAL_1:;
}

bool check(int a, int b, int c, int x) {
 GOAL_2:;
 return (a*x*x + b*x + c == 0);
}

int main() {
 GOAL_0:;
 int a = __VERIFIER_nondet_int();
 int b = __VERIFIER_nondet_int();
 int c = __VERIFIER_nondet_int();
 if(b*b >= 4*a*c) {
  GOAL_4:;
  while(1) {
   GOAL_6:;
   int x = __VERIFIER_nondet_int();
   if(x <= 0 || x > 100) {
    GOAL_8:;
    reach_error();
   }
   else {
    GOAL_9:;
   }
   if(check(a, b, c, x)) {
    GOAL_10:;
    return 0;
   }
   else {
    GOAL_11:;
   }
  }
  GOAL_7:;
 }
 else {
   GOAL_5:;
   reach_error();
 }    
 GOAL_3:;
 return 0;
}
        \end{lstlisting}
        \caption{}
        \label{figure:InstrumentedCode}
    \end{subfigure} %
    \begin{subfigure}{0.29\textwidth}
        \centering
        \includegraphics[width=\textwidth]{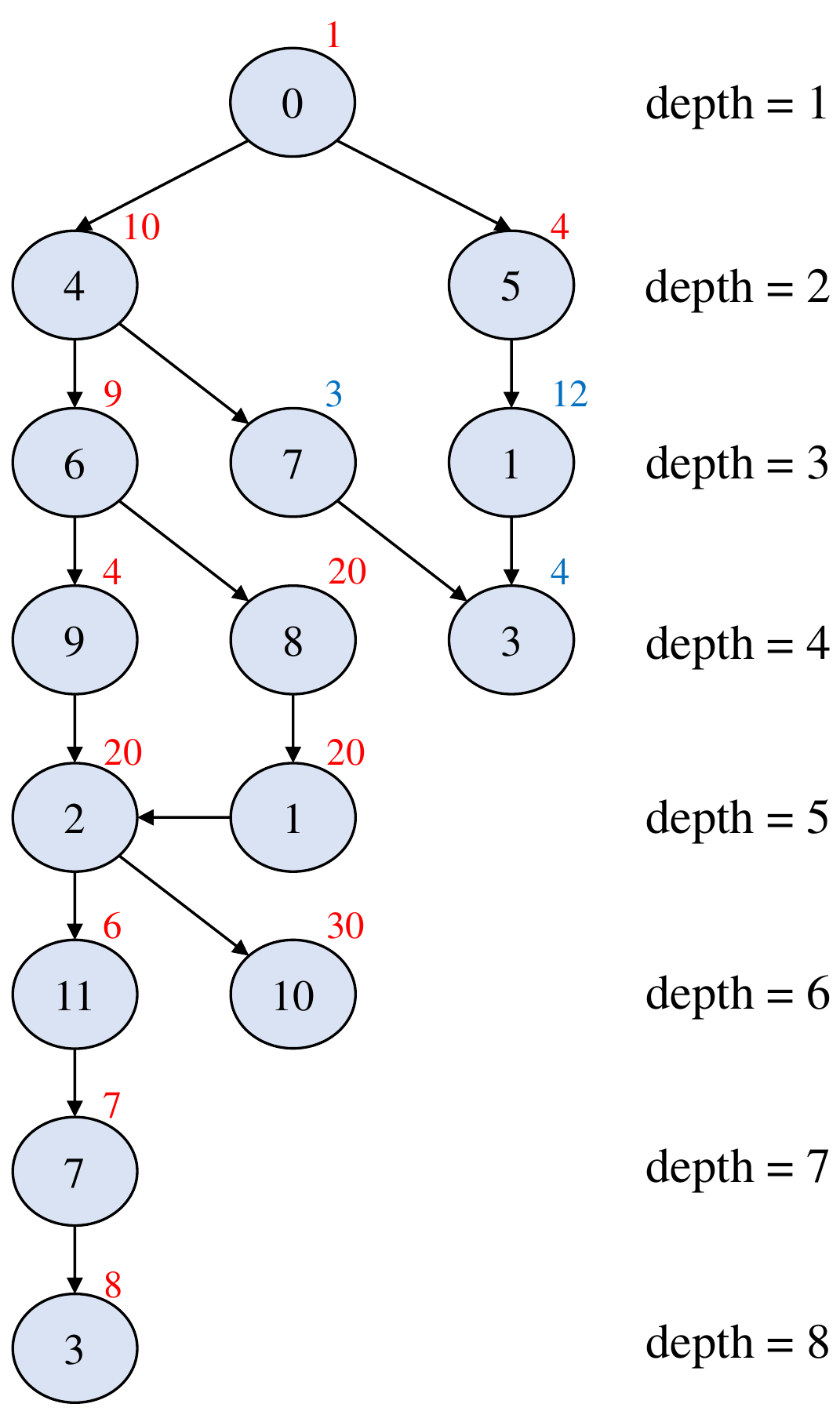}
        \caption{}
        \label{figure:ReachGraphRunning}
    \end{subfigure}
    \caption{\textcolor{black}{An example of a) a C program, b) the corresponding instrumented code, and c) the resulting {goals tree, their depth in the code, and resulting rank values}}.}
    \label{figure:IllustrativeExample}
\end{figure}

In Figure \ref{figure:IllustrativeExample} we introduce a short C program which we use as a running example to demonstrate the main code transformations throughout this section. The presented program accepts coefficients of a quadratic polynomial and an integer candidate solution in the range [1,100] as input from the user. It terminates successfully if the provided candidate solves the equation. However, the program returns an error if the given equation does not have real solutions or the input candidate value is outside the [1,100] range.

\begin{algorithm}[ht!]
\SetAlgoLined
\begin{algorithmic}[1]
\STATE {$P:= get\_input\_PUT()$}
\tcp*{{Code Instrumentation}}
\STATE{$P' := inject\_goal\_labels(P)$} \label{FuSeAlgo:inject}
\STATE{$G := get\_list\_of\_sorted\_goals(P')$\;} \label{FuSeAlgo:sort}

\STATE{$T := \emptyset; B := \emptyset$}\tcp*{{Seed Generation}}
\STATE{$\{S,G_{cov},T\} := generate\_seeds(P') $} \label{FuSeAlgo:SeedGen}
\STATE{$G.remove\_goals(G_{cov})$\;}
\WHILE{$G \not= \emptyset$ {\bf or} $timeout$} \label{FuSeAlgo:TG_Start}  
    \STATE{$g := G.pop()$} { \tcp*{{Start of main loop}}}
    \STATE{$\{output, G_{cov}\} := run\_fuzzer(P', g, S, fuzz_{timeout})$\;}    
    \IF{$G_{cov} \not = \emptyset$}
        \STATE{{$\{T,G,S\} := run\_tracer(P', output,T,G,G_{cov},S)$}}
    \ENDIF \\
    \IF{$g \in G_{cov}$}
        \STATE{{\bf continue}\;}
    \ENDIF \\
    \STATE{$\{output, res\} := run\_bmc(P', g, bmc_{timeout})$\;} \label{FuSeAlgo:bmc_start}
    \IF{$res = success$}
        \STATE{$\{T,G,S\} := run\_tracer(P', output, T,G,\emptyset,S)$}
    \ELSE
        \IF{$output \not = \emptyset$}
            \STATE{$B := B \cup generate\_bug\_report(P', output)$\;}
        \ELSE
            \STATE{{$\{T,G,S\} := run\_tracer(P', output,T,G,\emptyset,S)$}}
        \ENDIF
    \ENDIF\label{FuSeAlgo:bmc_end}
\ENDWHILE{\tcp*{{End of main loop}}}
\IF{{$G\not = \emptyset$}}
\STATE{{$\{testcases,G_{cov}\} := run\_selective\_fuzzer(P')$}}
\STATE{{$T := T \cup testcases$}}
\ENDIF \label{FuSeAlgo:TG_end}
\RETURN{$\{T, B, G\}$\;}
\end{algorithmic}
\caption{{\em FuSeBMC} algorithm}
\label{alg:FuSeBMC}
\end{algorithm}

\subsection{{Code Instrumentation \& Static Analysis}} 
{At this stage, \fb{} instruments the PUT and performs multiple static analyses. It takes the PUT (i.e., a C program) and a property file as inputs and produces three files: the instrumented program, \textit{Goal Queue}, and \textit{Consumed Input Size}.}

\subsubsection{{Code Instrumentation}}
\label{sec:CodeInstrumentation}

\textit{FuSeBMC} uses \textcolor{black}{the} Clang tooling infrastructure~\cite{CLANG} at its front-end to parse the input C program and traverse the resulting Abstract Syntax Tree (AST), recursively injecting goal labels into {the PUT}. 
{This process is guided by the \textit{FuSeBMC} code coverage criteria. Namely, {\em FuSeBMC} inserts labels 
inside} {conditional statements, loops, and functions as follows.}
\begin{itemize}
    \item For conditional statements: the label is inserted at the beginning of the block whether the statement is an \texttt{if}, \texttt{else}, or an instrumented empty \texttt{else}.
    
    \item For loops: the label is placed at the beginning of the loop body and right after exiting the loop. 

    \item For functions: labels are injected at the beginning and at the end of the function body.
\end{itemize}

{Furthermore, \textit{FuSeBMC} adds declarations for several standard C library functions, \textcolor{black}{such as \textit{``printf, ``strcpy'', ``memset''} and other C language functions}, to ensure that we cover the majority of the functions that we may encounter in large programs while also maintaining the proper operation of our approach.} \textcolor{black}{The resulting \emph{instrumented code} that has the labels injected is functionally equivalent to the original C program.} Figure~\ref{figure:InstrumentedCode} demonstrates an example of the described code instrumentation for the program in Figure~\ref{figure:original_code}.

\subsubsection{{Static Analysis}}
\label{sec:GoalsRanking}

Apart from the required code instrumentation, Clang produces compilation error and warning messages and utilizes its static analyzer to simplify the input program (e.g., calculating the sizes of expressions, evaluating static asserts, and performing constants propagation) \cite{LLVM:CGO04}.

\textcolor{black}{{Furthermore, the Clang static code analyzer produces the \emph{Consumed Input Size}, \textcolor{black}{which represents the minimum number of bytes in the input stream required for fuzzing.}} This information plays an important role in enhancing the fuzzing process (see Section \ref{sec:Fuzzers}).}

{Another function of static analysis is to identify the ranges of the input variables. This is performed by collecting branch conditions that match a pattern $(x \circ val)$, where $x$ is the name of the variable, $val$ is a numeric value, and $\circ \in \{>, \ge, <, \le, =, \not=\}$. Both fuzzers use this information to generate inputs only from the identified ranges.}

{Finally,} {\em FuSeBMC} analyzes the instrumented {code and ranks the injected goal labels.} {Each} goal label {is attributed} with its origin information (i.e.,\ if statement, while loop, end of function) and its depth in the instrumented program.
Then {\em FuSeBMC} sorts all goals using one of the two strategies (line \ref{FuSeAlgo:sort} of Algorithm \ref{alg:FuSeBMC}): (1) based on their depth (i.e., {depth-first search}), or (2) based on their {\em rank} scores calculated as a product of a goal's depth and its {\em power} score - a value between 1 and 5 describing the goal's branching power. Each power score was decided upon via experimental analysis. \textcolor{black}{The {\em if} statement goals are assigned a score of 5 (goals 4, 8 and 10 in Figure~\ref{figure:InstrumentedCode}), the {\em function} goals - 4 (goals 1 and 2 in Figure~\ref{figure:InstrumentedCode}, note that main function goals are scored differently), the {\em loop} goals - 3 (goal 6 in Figure~\ref{figure:InstrumentedCode}) and the {\em else} goals - 2 (goal 5 in Figure~\ref{figure:InstrumentedCode}). All remaining types of goals (i.e., {\em end-of-main} (goal 3 in Figure~\ref{figure:InstrumentedCode}), {\em empty-else} (goals 9, 11 in Figure~\ref{figure:InstrumentedCode}), and {\em after-loop} goals (goal 7 in Figure~\ref{figure:InstrumentedCode})) are assigned a value of 1.}

In general, the goal sorting improves overall {\em FuSeBMC} performance. \textcolor{black}{Using the depth-first strategy, {{\em FuSeBMC} attempts to cover the deeper goals first.}} This is beneficial since all preceding goals on the path to a deep goal can be ignored during subsequent fuzzing as the same \textcolor{black}{test-case} covers them. On the other hand, the ranking strategy allows for the prioritization of conditional branches as they may lead to multiple goals that increase potential code coverage. 

Figure~\ref{figure:ReachGraphRunning} features the resulting {goals tree} for the instrumented code from Figure~\ref{figure:InstrumentedCode} {(with \texttt{GOAL\_0} representing the entry point of the program, i.e., the \texttt{main} function)}.
Note that {\em FuSeBMC} builds {it} based on the original Clang AST without analyzing the code for trivially unreachable goals. For example, labels \texttt{GOAL\_7} and \texttt{GOAL\_3} can never be reached during the program's execution. However, this will not be reflected in the goals tree.

The goal's depth value is assigned at its highest depth. Therefore, labels \texttt{GOAL\_1}, \texttt{GOAL\_7}, and \texttt{GOAL\_3} are assigned depth values of 5, 7, and 8, respectively. When the first ranking strategy is applied, two goals at the same depth are ordered in the ascending order of their label names. Using the second-ranking strategy, two goals with the same {\em rank} value are processed in the ``power score first'' manner. For example, {\texttt{GOAL\_8}} will be placed in front of {\texttt{GOAL\_1}} and {\texttt{GOAL\_2}} since it has a higher power score (5 vs 4). Hence, {\em FuSeBMC} will process the goal labels in the following orders \{3,7,10,11,1,2,8,9,6,4,5\}, \{10,8,1,2,4,6,3,7,11,5,9\} using the first and second sorting strategies, respectively. 
{Finally, the list of goals is stored in the shared memory as the \emph{Goal Queue}. This queue can be modified by the BMC and Tracer engines during the consecutive stages to remove the goal labels that have been covered.}

\subsubsection{{Shared Memory}}
\label{sec:SharedMemory}

{The set of data files that each component of \textit{FuSeBMC} has access to (both for reading and writing) is called \textit{Shared Memory}. Apart from \textit{Instrumented Code}, \textit{Consumed Input Size} and \textit{Goal Queue} discussed above, it contains \textit{Seeds Store} -- a collection of seeds used by the fuzzer for test generation, \textit{\textcolor{black}{test-cases}} -- all \textcolor{black}{test-cases} generated by \textit{FuSeBMC}, and \textit{Goal Covered Array} -- list of all goal labels that have been covered by the produced \textcolor{black}{test-cases}.}

\subsection{Seed Generation}
\label{sec:SeedGeneration}

\begin{algorithm}[ht!]
\SetAlgoLined
\hspace*{\algorithmicindent} \textbf{Input}: $P' : \text{instrumented file}$  \\
    \hspace*{\algorithmicindent} \textbf{Output}: \textit{$S$} : \text{set of seeds}, \textit{$G_{cov}$} : goals covered, $T$: generated \textcolor{black}{test-cases}
\begin{algorithmic}[1]
\STATE{$S := generate\_primary\_seeds(P')$} \label{SeedGen:l1} 
\STATE{$\{\textcolor{black}{test-cases}, G_{cov}\} := run\_fuzzer(P', S, fuzz_{timeout})$} \label{SeedGen:l2}
\IF{$G_{cov} \not = \emptyset$}
        \STATE{$T := T \cup testcases$\;}
        \STATE{$S := S \cup generate\_seed(testcases)$}
    \ENDIF\label{SeedGen:l3} \\
\STATE{$\{output, res\} := run\_light\_bmc(P', bmc_{timeout})$}
\IF{$res = success$}
        \STATE{$S := {S\cup{output}}$\;}
        \STATE{$T := T \cup generate\_testcases(P', \{output\})$\;}
    \ELSE
        \IF{$output = \emptyset$}
            \STATE{$T := T \cup generate\_testcases(P', S)$\;}
        \ENDIF
    \ENDIF
    \RETURN{$\{S, G_{cov}, T\}$\;}
\end{algorithmic}
\caption{{$generate\_seeds()$ algorithm}}
\label{alg:seedGen}
\end{algorithm}

\textcolor{black}{Having ranked the goals, {\em FuSeBMC} carries out seed generation (line \ref{FuSeAlgo:SeedGen} of Algorithm \ref{alg:FuSeBMC} {and Algorithm \ref{alg:seedGen} where it is described in detail}) as a preliminary step before full coverage analysis (i.e., {\em test generation}) begins. In this phase, {\em FuSeBMC} simplifies the target program by limiting loop bounds, {and utilizes} the information about the input ranges. {Then \textit{FuSeBMC} applies the fuzzer and the BMC engine (for 5 and 15 seconds, respectively) for a short time in succession.}}

\textcolor{black}{{Since the seed store is empty at this point, \textit{FuSeBMC} performs \textit{primary seed generation} (Line \ref{SeedGen:l1} Algorithim \ref{alg:seedGen}) to enable the fuzzing process. This procedure involves generating} binary seeds (i.e., a stream of bytes) {based on \textit{Consumed Input Size} and the input constraints collected during static analysis}. {In detail, it generates three sequences of bytes, where 1) all bytes have a value of 0, 2) all bytes have a value of 1, and 3) all byte values are drawn randomly from the identified input ranges. Then the fuzzer is initialized with the primary seeds and} is run for a short time to produce \textcolor{black}{test-cases} that are then converted into new seeds and added to the \textit{Seeds store} (\textcolor{black}{see Figure~\ref{figure:FuSeBMCFramework} and lines \ref{SeedGen:l2}---\ref{SeedGen:l3} in Algorithm \ref{alg:seedGen}).}}

\textcolor{black}{{When the seed generation by fuzzing is finished, \textit{FuSeBMC} executes the BMC engine for each goal label in the \textit{Goal Queue}. To minimize the execution time, it is run with "lighter" settings: all implicit checks (i.e., memory safety, arithmetic overflows) and assertion checks are disabled, and the bound for loop unwinding is reduced. If a goal label is reached successfully, the BMC engine produces a witness -- a set of program inputs that lead to that goal label. The sequence of these input values is added as a seed to the \textit{Seed Store}.}}

\textcolor{black}{{All new seeds produced by the fuzzer and the BMC engine} are {deemed} {\em smart} due to their powerful effect on code coverage. Conceptually, bounded model checkers use SMT solvers to produce \textcolor{black}{test-cases} that {resolve} complex {branch conditions (i.e., guards)}. Such guards (for example, lines 5 and 12 in Figure~\ref{figure:original_code}) pose a challenge to a fuzzer~\cite{stephens2016driller} as it relies on mutating the given seed randomly and is therefore unlikely to satisfy the {branch} condition. Seeds produced by BMC {help solve} this issue since they can be passed to a fuzzer, which can then advance deeper behind the complex guards into the target program (which is usually hard for a bounded model checker).}

\subsection{Test Generation}
\label{sec:TestGeneration}

Following seed generation, {\em FuSeBMC} begins the main coverage analysis phase (\textcolor{black}{lines \ref{FuSeAlgo:TG_Start}---\ref{FuSeAlgo:TG_end} of Algorithm \ref{alg:FuSeBMC})}. {\em FuSeBMC} incorporates three engines to carry out this analysis: two fuzzers {(main fuzzer and selective fuzzer)} and a bounded model checker. {Here, the main fuzzer and the BMC engine are} run with longer timeouts than during the seed generation stage. {Briefly, the fuzzers utilize the \textit{smart} seeds produced at the previous stage, and} generate \textcolor{black}{test-cases} by randomly mutating the program's input and running it to analyze code coverage. The bounded model checker determines the reachability of particular goal labels {similarly to the \textit{seed generation} stage}. \textcolor{black}{{\em FuSeBMC}'s \tracer{} component {aids} the above engines {by replaying a light-weight analysis of the produced \textcolor{black}{test-cases} and updating the} \emph{Shared Memory}.}
In the following subsections, we discuss the {\em FuSeBMC} components involved in coverage analysis in greater detail.

\subsubsection{Main Fuzzer}
\label{sec:Fuzzers}

In \textit{FuSeBMC} {we implement} a modified version of the American Fuzzy Lop (AFL) tool~\cite{americanfuzzylop_2021}. The modified AFL generates \textcolor{black}{test-cases} based on the evolutionary algorithm implemented in AFL~\cite{bohme2017directed}. 

The standard algorithm implemented in the AFL tool works as follows. Firstly, an initial fixed-size input stream is generated using the provided seed (a random seed is used if not explicitly specified). Secondly, the target program is repeatedly executed with the randomly mutated input. If the target program does not reach any new states after multiple input mutation rounds, a new byte is added to {or removed from} the input stream, and the mutation process restarts. The above algorithm continues until an internal timeout is reached or the fuzzer finds inputs that fully cover the program. In general, the AFL's mutation algorithm relies heavily on the quality of the initial seeds for higher code coverage. Therefore, generating seeds with higher coverage potential is crucial.

{\em FuSeBMC} modifies the original AFL fuzzer as follows.
\begin{enumerate}
    \item {It performs additional instrumentation to the PUT} to minimize its execution overhead by limiting the bounds of loops heuristically identified as potentially infinite. Note that these bounds can be iteratively changed between the AFL runs.

    \item \textcolor{black}{The mutation operators are modified to generate only inputs from the ranges identified during the {static} code analysis.}

    \item {It controls the size of the generated \textcolor{black}{test-cases} via the {\em Consumed Input Size}. In detail, the minimum size of the \textcolor{black}{test-cases} produced by the fuzzer is set to the current value of the consumed input size. This allows counter-acting the size selection bias of the AFL mutation algorithms, which tend to favor a reduction of the number of bytes in the generated \textcolor{black}{test-cases} (instead of adding extra bytes) between the mutation rounds. At the same time, the modified fuzzer can control the maximum size of the produced \textcolor{black}{test-cases}. For example, when the \textit{Consumed Input Size} starts growing gradually during the fuzzing process (a behaviour often observed in programs accepting input in an infinite loop), the maximum test-case size is set to prevent performance degradation.}

    \item {It outputs the list of goals covered by the produced \textcolor{black}{test-cases} and records them in \textit{Goals Covered Array}.}
\end{enumerate}

\subsubsection{Bounded Model Checker}
\label{sec:BMC}

{\textit{FuSeBMC} uses \textit{ESBMC}} to check for the reachability of a given goal label within the instrumented program (\textcolor{black}{lines \ref{FuSeAlgo:bmc_start}---\ref{FuSeAlgo:bmc_end} of Algorithm \ref{alg:FuSeBMC}).} If it concludes that the current goal is reachable {it produces a} counterexample {that} can be turned into a witness -- a sequence of inputs that leads the program's execution to that goal label -- which is then used to generate a \textcolor{black}{test-case}. Every new \textcolor{black}{test-case} thus discovered is also added to the {\em Seed Store} to be used by the fuzzers. Even if the BMC runs out of time or memory, its progress in reducing the input ranges is saved {as an \textit{incomplete seed} -- a sequence of input values that lead the PUT execution part of the way towards the given goal label}.


\subsubsection{Tracer}
\label{sec:Tracer}

\begin{algorithm}[ht!]
\SetAlgoLined
\hspace*{\algorithmicindent} \textbf{Input}: $P': \text{instrumented file}, \textit{output}: \text{engines output}$, \\$T$: generated \textcolor{black}{test-cases}, $G$: list of goals, $G_{cov}$: covered goals list, \textit{$S$}: \text{set of seeds} \\
    \hspace*{\algorithmicindent} \textbf{Output}: $T$:generated \textcolor{black}{test-cases}, \textit{$G$}: list of goals, \textit{$S$}: \text{set of seeds} 
\begin{algorithmic}[1]
\IF{$output$ from BMC and incomplete}
    \STATE{$testcase := complete\_testcase(output)$}

\ELSE
\STATE{$testcase:= generate\_\textcolor{black}{test-cases}(output,G_{cov})$}
\ENDIF
\STATE{$T:= T \cup testcase$\;}
\STATE{$G_{cov}:= run\_testcase(P', testcase)$}
\STATE{$G.remove\_goals(G_{cov})$\;}
\STATE{$S:= S\cup{testcase}$\;}
\RETURN $\{T,G,S\}$
\end{algorithmic}
\caption{{$run\_tracer()$ algorithm}}
\label{alg:Tracer}
\end{algorithm}

The {\em Tracer} subsystem determines the goals covered by \textcolor{black}{test-cases} produced by the bounded model checker {and the fuzzer}. {Whenever a \textcolor{black}{test-case} is produced,} \textit{Tracer} compiles the instrumented program together with the {newly generated} \textcolor{black}{test-cases} and runs the resulting executable. {Before the compilation, it performs additional instrumentation to the \textcolor{black}{test-case} to output} information about the {PUT} input size, the types of input variables, and the visited goals. This information is dynamically updated in the {\em Shared Memory} (i.e., {\em Goals Covered Array} and {\em Consumed Input Size}).

The Tracer also analyses the \textcolor{black}{test-cases} produced by the other two engines to add the highest impact cases (i.e., the \textcolor{black}{test-cases} leading to new goals or reaching the maximum analysis depth) to the {\em Seeds Store}.

{Another responsibility of \tracer{} is to handle the partial output of the bounded model checker when it reaches the timeout, outputting an incomplete counterexample. \tracer{} completes such counterexamples randomly and performs the coverage analysis and updates \textit{Seeds Store} as described above.}


\subsubsection{{Selective Fuzzer}}

{The selective fuzzer's \cite{alshmrany2021fusebmc} main function is to attempt to reach the remaining uncovered goals after the iterative process of applying the fuzzer, and the BMC engine has finished. Similarly to the main fuzzer, it utilizes information about the identified ranges of the input variables to produce inputs for the PUT. At the same time, it implements a complementary test generation approach. It produces random values from the given input ranges -- in contrast to the mutation-based approach used in the main fuzzer. The selective fuzzer terminates upon covering all the remaining goals or reaching the timeout.}

\section{Evaluation}
\label{sec:ExperimentalEvaluation}

\subsection{Description of Benchmarks and Setup}
\label{sec:BenchmarksAndSetup}

{To assess the performance of \textit{FuSeBMC v4}, we evaluated its participation in Test-Comp 2022 \cite{TESTCOMP22}, and also compared it to the results obtained by the previous version of the tool, \textit{FuSeBMC v3}, in Test-Comp 2021 \cite{TESTCOMP21}.}

{Test-Comp is a software testing competition where the participating tools compete in automated \textcolor{black}{test-case} generation. All \textcolor{black}{test-case} generation tasks in Test-Comp are divided into two categories: \textit{Cover-Branches} and \textit{Cover-Error}. The former requires producing a set of \textcolor{black}{test-cases} that maximize \textit{code coverage} (in particular, \textit{branch coverage}) for the given C program. The latter deals with \textit{error coverage}: generating a \textcolor{black}{test-case} that leads to the predefined error location (i.e., explicitly marked error function) within the given C program. In \textit{Cover-Branches},} code coverage is measured by the TestCov~\cite{beyer2019testcov} tool, which assigns a score between 0 and 1 for each task. {For example,} if a competing tool achieves $80$\% code coverage on a particular task, it is assigned a score of 0.8 for that task and so forth. Overall scores for the subcategories are calculated by summing the individual scores for each task in the subcategory and rounding the result.
{In \textit{Cover-Error},} each tool earns a score of 1 if it can provide a \textcolor{black}{test-case} that reaches the error function and gets a 0 score otherwise.
{Each category is further divided into multiple subcategories (see Tables \ref{table:ImprovementsCoverBranches} and \ref{table:ImprovementsCoverError}) based on the most prominent program features and/or the program's origin.}
Most programs in Test-Comp are taken from SV-COMP~\cite{beyer2software} -- the largest and most diverse open-source repository of software verification tasks. {It contains hand-crafted and real-world} C programs with loops, arrays, bit-vectors, floating-point numbers, dynamic memory allocation, and recursive functions, event-condition-action software, concurrent programs, and BusyBox\footnote{https://busybox.net/} software.

{Both Test-Comp 2021 and Test-Comp 2022 evaluations} were conducted on {servers} featuring an 8-core (4 physical cores) Intel Xeon E3-1230 v5 CPU @ 3.4 GHz, 33 GB of RAM and running x86-64 Ubuntu 20.04 with Linux kernel 5.4. Each test suite generation run was limited to 8 CPU cores, 15 GB of RAM, and 15 minutes of CPU time, while each test suite validation run was limited to 2 CPU cores, 7 GB of RAM, and 5 minutes of CPU time. \ahmedhappy{In 2021, \fb{} distributed its allocated time to its various engines as follows. \textcolor{black}{The fuzzer received 150s as a time limit when running on benchmarks from the \textit{Cover-Error} category and 70s on benchmarks from the \textit{Cover-Branches} category.} The bounded model checker received 700s and 780s on benchmarks from the two categories, respectively. Finally, the selective fuzzer received 50s for benchmarks from both categories. In 2022, these figures were tweaked. The seed generation received 20s for benchmarks from both categories. The fuzzer received 200s and 250s on benchmarks from \textit{Cover-Error} and \textit{Cover-Branches}, respectively, the bounded model checker 650s and 600s, and the allocation for the selective fuzzer was decreased from 50s to 30s from the previous year.}


{Although the hardware setup remained unchanged across the two competition editions, the set of test generation tasks was significantly expanded. Namely, the} task set {in Test-Comp 2021} consisted of {3173 tasks: 607 in the \textit{Cover-Error} category, and 2566 in the \textit{Cover-Branches} category. By contrast, Test-Comp 2022 was expanded to contain}
4236 test tasks: 776 in the \textit{Cover-Error} category and 3460 in the \textit{Cover-Branches} category {(including a new subcategory \textit{ProductLines} introduced into both categories)}. {We have considered this when discussing the performance of two versions of \textit{FuSeBMC} in Section \ref{sec:FuSeBMCImprovement}. A detailed report of the results produced by the competing tools in both Test-Comp 2021\footnote{\url{https://test-comp.sosy-lab.org/2022/results/results-verified/}} and Test-Comp 2022\footnote{\url{https://test-comp.sosy-lab.org/2021/results/results-verified/}} is available online.}

\textit{FuSeBMC} source code is written in C++ and Python; it is available for download from GitHub\footnote{\url{https://github.com/kaled-alshmrany/FuSeBMC}}. The latest release of \textit{FuSeBMC} is v4.1.14. \textit{FuSeBMC} is publicly available under the terms of the MIT license. Instructions for building \textit{FuSeBMC} from the source code are given in the file \textit{README.md}.

\subsection{Objectives}
\label{sec:Objectives}

{The main goal of our experimental evaluation is to assess the improvements of \textit{FuSeBMC v4} and its suitability for achieving high code coverage and error coverage in open-source C programs. As a result, we identify three key evaluation objectives:}

\begin{tcolorbox}
    \begin{enumerate}
        \item[{\textbf{O1}}] {\textbf{(Performance
        Improvement)} Demonstrate that \textit{FuSeBMC v4} outperforms 
        \textit{FuSeBMC v3} in both code coverage and 
        error coverage.}

        \item[{\textbf{O2}}] \textbf{(Coverage Capacity)} {Demonstrate that} \textit{FuSeBMC v4} achieves higher code coverage for C programs than other state-of-the-art software testing tools.
                
        \item[{\textbf{O3}}] {\textbf{(Error Detection)} Demonstrate that} \textit{FuSeBMC v4} finds more errors in C programs than other state-of-the-art software testing tools.
    \end{enumerate}
\end{tcolorbox}

\subsection{Results}
\label{sec:Results}

\subsubsection{{\textit{FuSeBMC v4} vs \textit{FuSeBMC v3}}}
\label{sec:FuSeBMCImprovement}
{Tables \ref{table:ImprovementsCoverBranches} and \ref{table:ImprovementsCoverError} contain the comparison of the \textit{FuSeBMC v4} and \textit{FuSeBMC v3} performances in \textit{Cover-Branches} and \textit{Cover-Error} categories of Test-Comp, respectively. \textit{FuSeBMC v3} achieved first place in \textit{Cover-Error}, fourth place in \textit{Cover-Branches}, and placed second overall in Test-Comp 2021, while \textit{FuSeBMC v4} reached first place in both categories and overall in Test-Comp 2022. However, considering the test generation task set has been significantly expanded in Test-Comp 2022, we analyze their relative performances in each subcategory. Namely, in \textit{Cover-Branches}, we compare average code branch coverage. In \textit{Cover-Error}, we compare the percentages of successfully detected errors demonstrated by both tools in every subcategory and the entire category (as well the improvements of \textit{FuseBMC v4} in comparison to \textit{FuSeBMC v3}), respectively.}

\begin{table*}
    \centering
    \caption{{Comparison of the average coverage (per subcategory and the category overall) achieved by \textit{FuSeBMC v4} and \textit{FuSeBMC v3} in the \textit{Cover-Branches} category in TestComp-2022 and TestComp-2021, respectively.}}
    \begin{tabular}{|l||c|c|c|}
        \hline
        \multirow{2}{*}{Subcategory} & \multicolumn{2}{c|}{{\% average coverage}} & Improvement \\
        \cline{2-3}
        & \textit{FuSeBMC v4} & \textit{FuSeBMC v3}& $\Delta\%$ \\
        \hline
        Arrays           & 82\%  & 71\%  & 11\% \\
        BitVectors       & 80\%  & 60\%  & 20\% \\
        ControlFlow      & 64\%  & 22\%  & 42\% \\
        ECA              & 37\%  & 17\%  & 20\% \\
        Floats           & 54\%  & 46\%  & 8\%  \\
        Heap             & 73\%  & 62\%  & 11\% \\
        Loops            & 81\%  & 71\%  & 10\% \\
        ProductLines     & 29\%  & {-}   & {-}  \\
        Recursive        & 85\%  & 68\%  & 18\% \\
        Sequentialized   & 87\%  & 76\%  & 11\% \\
        XCSP             & 90\%  & 82\%  & 8\%  \\
        Combinations     & 61\%  & 7\%   & 53\% \\
        BusyBox          & 34\%  & 1\%   & 32\% \\
        DeviceDrivers    & 20\%  & 12\%  & 8\%  \\
        SQLite-MemSafety & 4\%   & 0\%   & 4\%  \\
        Termination      & 92\%  & 87\%  & 5\%  \\
        \hline
        {\textit{Cover-Branches}}    & 61\%  & 45\%  & 16\% \\
        \hline
    \end{tabular}
    \label{table:ImprovementsCoverBranches}
\end{table*}

{Table \ref{table:ImprovementsCoverBranches} shows that \textit{FuSeBMC v4} advanced in each subcategory, including the overall average improvement of 16\%} in the \textit{Cover-Branches} category in comparison to \textit{FuSeBMC v3}. {The greatest increase (i.e., 53\%) was demonstrated in the \textit{Combinations} subcategory. \textit{FuSeBMC v3} achieved eighth place in this subcategory in Test-Comp 2021, while \textit{FuSeBMC v4} reached first place in \textit{Combinations} in Test-Comp 2022. We attribute this success to the modifications in the seed generation phase of \textit{FuSeBMC v4} (in particular, the introduction of \textit{smart seeds}). Table~\ref{table:FuSeBMCVersionsComparison} presents a subset of generation tasks from the \textit{Combinations} subcategory where \textit{FuSeBMC v4} demonstrated the most striking improvement. 
It can be seen that \textit{FuSeBMC v3} provided very low code coverage of $\sim 6.52\%$ for these tasks on average, while \textit{FuSeBMC v4} increased this number to $\sim 90.14\%$ (i.e., 83.62\% average improvement).}


\begin{table*}
    \centering
    \caption{Comparison of code coverage achieved by \textit{FuSeBMC v4} and \textit{FuSeBMC v3} in a subset of tasks from the \textit{Combinations} subcategory.}
    \begin{tabular}{|l||c|c|c|}
    \hline
    & \multicolumn{2}{c|}{\% coverage} & Improvement\\
    \cline{2-3}
    Task name & \textit{FuSeBMC v4} & \textit{FuSeBMC v3} & $\Delta\%$ \\ 
    \hline
    \texttt{pals\_lcr.3.1.ufo.BOUNDED-6.pals+Problem12\_label01.yml}        & $94.90$\% & $13.30$\% & $81.60$\% \\ 
    \texttt{pals\_lcr.3.1.ufo.UNBOUNDED.pals+Problem12\_label02.yml}        & $84.40$\% & $5.19$\% &  $79.21$\% \\ 
    \texttt{pals\_lcr.4.1.ufo.BOUNDED-8.pals+Problem12\_label04.yml}        & $94.10$\% & $4.44$\% &  $89.66$\% \\ 
    \texttt{pals\_lcr.4\_overflow.ufo.UNBOUNDED.pals+Problem12\_label05.yml}& $94.00$\% & $11.50$\% & $82.50$\% \\ 
    \texttt{pals\_lcr.5.1.ufo.UNBOUNDED.pals+Problem12\_label05.yml}        & $86.20$\% & $0.78$\% &  $85.42$\% \\
    \texttt{pals\_lcr.5\_overflow.ufo.UNBOUNDED.pals+Problem12\_label09.yml}& $94.00$\% & $4.82$\% &  $89.18$\% \\ 
    \texttt{pals\_lcr.6.1.ufo.BOUNDED-12.pals+Problem12\_label09.yml}       & $92.90$\% & $5.18$\% & $87.72$\%  \\ 
    \texttt{pals\_lcr.7\_overflow.ufo.UNBOUNDED.pals+Problem12\_label09.yml}& $92.60$\% & $5.31$\% &  $87.29$\% \\ 
    \texttt{pals\_lcr.8.ufo.UNBOUNDED.pals+Problem12\_label08.yml}          & $78.20$\% & $8.17$\% &  $70.03$\% \\ \hline
    Average value & 90.14\% & 6.52\% & 83.62\% \\
    \hline
    \end{tabular}
    \label{table:FuSeBMCVersionsComparison}
\end{table*}

{As for the \textit{Cover-Error} category,} \textit{FuSeBMC v4} progressed by $14\%$ on average in comparison to \textit{FuSeBMC v3} (see Table~\ref{table:ImprovementsCoverError}). {\textit{FuSeBMC v4} improved in the majority of subcategories while showing no change in three subcategories: both \textit{FuSeBMC} versions achieved the highest possible result of 100\% in \textit{BitVectors}, \textit{FuSeBMC v4} failed to advance the number of detected errors past 95\% in \textit{Recursive}, while \textit{FuSeBMC v4} could not identify any errors in \textit{DeviceDrivers} similarly to \textit{FuSeBMC v3}. \textcolor{black}{Also, \textit{FuSeBMC} v4 demonstrated a performance degradation of $2$\% in the \textit{XCSP} subcategory. This phenomenon is attributable to certain programs that redefine C library functions and encompass numerous conditional statements, resulting in a slowdown in seed generation and excessive utilization of the tool's resources.}}

\begin{table*}
    \centering
    \caption{{Comparison of the percentages of the successfully detected errors (per category and the category overall) by \textit{FuSeBMC v4} and \textit{FuSeBMC v3} in the \textit{Error Coverage} category in TestComp-2022 and TestComp-2021, respectively.}}
    \begin{tabular}{|l||c|c|c|}
        \hline
        \multirow{2}{*}{Subcategory} & \multicolumn{2}{c|}{{\% errors detected}} & Improvement \\
        \cline{2-3}
        & \textit{FuSeBMC v4} & \textit{FuSeBMC v3}& $\Delta\%$ \\
        \hline
        Arrays         & 99\%  & 93\%  & 6\%   \\
        BitVectors     & 100\% & 100\% & 0\%   \\
        ControlFlow    & 100\% & 25\%  & 75\%  \\
        ECA            & 72\%  & 44\%  & 28\%  \\
        Floats         & 100\% & 97\%  & 3\%   \\
        Heap           & 95\%  & 80\%  & 14\%  \\
        Loops          & 93\%  & 83\%  & 10\%  \\
        ProductLines   & 100\% & {-}   & {-} \\
        Recursive      & 95\%  & 95\%  & 0\%   \\
        Sequentialized & 95\%  & 94\%  & 1\%   \\
        XCSP           & 88\%  & 90\%  & -2\%  \\
        BusyBox        & 15\%  & 0\%   & 15\%  \\
        DeviceDrivers  & 0\%   & 0\%   & 0\%   \\
        \hline
        {\textit{Cover-Error}}  & 81\%  & 67\%  & 14\%  \\
        \hline
    \end{tabular}
    \label{table:ImprovementsCoverError}
\end{table*}

{
Additionally,} we compared the performance of \textit{FuSeBMC v4} utilizing smart seeds with the version of \textit{FuSeBMC v4} using only {primary} seeds (\ie all zeros, all ones, and randomly chosen values) {on the \textit{ECA} (which stands for \textit{event-condition-action} systems) subcategory in \textit{Cover-Error} (where \textit{FuSeBMC v4} demonstrated 28\% improvement in comparison to \textit{FuSeBMC v3} in the competition settings; see Table \ref{table:ImprovementsCoverError}). It contains 18 \textcolor{black}{test-case} generation tasks with C programs featuring input validation that involves relatively complex mathematical expressions. Such a program feature is notoriously difficult for the fuzzers whose initial seed is based on a random choice.} Table~\ref{table:SeedsComparison} presents the results obtained by the versions of \textit{FuSeBMC v4} with smart seeds and with primary seeds. It can be seen that smart seeds allow detecting $5$ more bugs than the version of \textit{FuSeBMC} using standard seeds.

\begin{table*}
    \centering
    \caption{Comparison of \textit{FuSeBMC v4} performance with smart seeds and with standard seeds, where TRUE shows that the bug has been detected successfully, UNKNOWN means otherwise.}
    \begin{tabular}{|l||c|c|}
    \hline
    & \multicolumn{2}{c|}{\textit{FuSeBMC v4}} \\
    \cline{2-3}
    Task name & Smart Seeds & {Primary} Seeds \\
    \hline
    \texttt{eca-rers2012/Problem05\_label00.yml}         &  TRUE   &  TRUE   \\
    \texttt{eca-rers2012/Problem06\_label00.yml}         &  TRUE   &  TRUE   \\
    \texttt{eca-rers2012/Problem11\_label00.yml}         &  TRUE   &  TRUE   \\
    \texttt{eca-rers2012/Problem12\_label00.yml}         &  TRUE   &  TRUE   \\
    \texttt{eca-rers2012/Problem15\_label00.yml}         &  TRUE   &  TRUE   \\
    \texttt{eca-rers2012/Problem16\_label00.yml}         &  TRUE   & UNKNOWN \\
    \texttt{eca-rers2012/Problem18\_label00.ymll}        &  TRUE   &  TRUE   \\
    \texttt{eca-rers2018/Problem10.yml}                  &  TRUE   &  TRUE   \\
    \texttt{eca-rers2018/Problem11.yml}                  &  TRUE   &  TRUE   \\
    \texttt{eca-rers2018/Problem12.yml}                  &  TRUE   & UNKNOWN \\
    \texttt{eca-rers2018/Problem13.yml}                  &  TRUE   & UNKNOWN \\
    \texttt{eca-rers2018/Problem14.yml}                  &  TRUE   & UNKNOWN \\
    \texttt{eca-rers2018/Problem15.yml}                  &  TRUE   & UNKNOWN \\
    \texttt{eca-rers2018/Problem16.yml}                  & UNKNOWN & UNKNOWN \\
    \texttt{eca-rers2018/Problem17.yml}                  & UNKNOWN & UNKNOWN \\
    \texttt{eca-rers2018/Problem18.yml}                  & UNKNOWN & UNKNOWN \\
    \texttt{eca-programs/Problem101\_label00.yml}        & UNKNOWN & UNKNOWN \\
    \texttt{eca-programs/Problem103\_label32.yml}        & UNKNOWN & UNKNOWN \\
    \hline
    \end{tabular}
    \label{table:SeedsComparison}
\end{table*}

{Overall, the results presented in Tables \ref{table:ImprovementsCoverBranches} and \ref{table:ImprovementsCoverError} provide sufficient evidence that the evaluation objective \textbf{O1} has been achieved.}

\subsubsection{{\textit{FuSeBMC v4} vs state-of-the-art}}
\textit{FuSeBMC v4} achieved {the overall} first place at Test-Comp 2022, obtaining a score of 3003 out of 4236 with the closest competitor, VeriFuzz~\cite{chowdhury2019verifuzz}, scoring 2971 and significantly outperforming several state-of-the-art tools such as LibKluzzer~\cite{le2020llvm}, KLEE~\cite{cadar2008klee}, CPAchecker~\cite{beyer2011cpachecker} and Symbiotic~\cite{Chalupa2021Symbiotic}
(see Table~\ref{table:Overall}).

\begin{table*}
    \centering
    \caption[Overall]{\textcolor{black}{{Test-Comp 2022 \textit{Overall} Results}{\footnotemark}. The table illustrates the scores obtained by all state-of-the-art tools overall, where we identify the best tool in bold.}}
    \begin{tabular}{|c||c|c|c|c|c|c|c|c|c|c|c|c|}
        \hline
        \multirow{2}{*}{} & \multicolumn{12}{c|}{Tool} \\
        \cline{2-13}
        \rotatebox[origin=c]{90}{Total \# tasks} &
        \rotatebox[origin=c]{90}{CMA-ES Fuzz} &
        \rotatebox[origin=c]{90}{CoVeriTest v2.0.1} &
        \rotatebox[origin=c]{90}{\textit{FuSeBMC} v4.1.14} &
        \rotatebox[origin=c]{90}{ HybridTiger v1.9.2 } &
        \rotatebox[origin=c]{90}{KLEE v2.2} &
        \rotatebox[origin=c]{90}{Legion v1.0} &
        \rotatebox[origin=c]{90}{Legion/SymCC} &
        \rotatebox[origin=c]{90}{LibKluzzer v1.0} &
        \rotatebox[origin=c]{90}{PRTest v2.2} &
        \rotatebox[origin=c]{90}{Symbiotic v9.0} &
        \rotatebox[origin=c]{90}{Tracer-X v1.2.0} &
        \rotatebox[origin=c]{90}{VeriFuzz v1.2.10} \\
        \hline
        4236 & 382 & 2293 & \textbf{3003} & 1830 & 2125 & 787 & - & 2658 & 945 & 2367 & 1069 & 2971 \\ \hline
    \end{tabular}
    \label{table:Overall}
\end{table*}
\footnotetext{\url{https://test-comp.sosy-lab.org/2022/results/results-verified/}}

{Table~\ref{table:CoverBranches} demonstrates the code coverage capabilities of \textit{FuSeBMC v4} in comparison to other state-of-the-art software testing tools. It can be seen that} \textit{FuSeBMC} achieved first place with an overall score of 2104 out of 3460. \textit{FuSeBMC} participated in {all} 16 subcategories, in 9 of which (\ie \textit{Arrays, BitVectors, Floats, Heap, Loops, ProductLines, Recursive, Combinations} and \textit{Termination}) it achieved first place and in 6 of which it reached second place. {The results presented in Table \ref{table:CoverBranches} allow us concluding that the evaluation objective \textbf{O2} has been achieved.}

\begin{table*}
    \centering
    \caption[Cover-Branches]{\textcolor{black}{{\textit{Cover-Branches} category results at \textit{Test-COMP 2022}}{\footnotemark}. The best score for each subcategory is highlighted in bold.}}
    \begin{tabular}{|l||c|c|c|c|c|c|c|c|c|c|c|c|c|}
        \hline
        \multirow{2}{*}{} & & \multicolumn{12}{c|}{Tool} \\
        \cline{3-14}
        \multirow{2}{*}{Subcategory} &
        \rotatebox[origin=c]{90}{Total \# tasks} &
        \rotatebox[origin=c]{90}{CMA-ES Fuzz} &
        \rotatebox[origin=c]{90}{CoVeriTest v2.0.1} &
        \rotatebox[origin=c]{90}{\textit{FuSeBMC} v4.1.14} &
        \rotatebox[origin=c]{90}{ HybridTiger v1.9.2 } &
        \rotatebox[origin=c]{90}{KLEE v2.2} &
        \rotatebox[origin=c]{90}{Legion v1.0} &
        \rotatebox[origin=c]{90}{Legion/SymCC~} &
        \rotatebox[origin=c]{90}{LibKluzzer v1.0} & 
        \rotatebox[origin=c]{90}{PRTest v2.2} &
        \rotatebox[origin=c]{90}{Symbiotic v9.0} &
        \rotatebox[origin=c]{90}{Tracer-X v1.2.0} &
        \rotatebox[origin=c]{90}{VeriFuzz v1.2.10} \\
        \hline
        Arrays                & 400  & 159 &    257    & \textbf{328} &     247    & 104  & 210  & 263  &    323     & 160 &     250   &    243    &      307  \\ 
        BitVectors            & 62   & 26  &\textbf{49}& \textbf{49}  &     16     & 33   & 34   & 45   &    48      & 33  &\textbf{49}&\textbf{49}&      48   \\ 
        ControlFlow           & 67   & 5   &    36     &      43      &     14     & 25   & 17   & 41   &    40      & 6   &     43    &    39     &\textbf{44}\\ 
        ECA                   & 29   & 0   &    6      &      11      &     2      & 7    & 3    & 3    &    10      & 2   &     10    &    8      &\textbf{12}\\ 
        Floats                & 226  & 53  &    113    & \textbf{122} &     92     & 19   & 72   & 62   &    103     & 46  &     55    &    56     &      119  \\ 
        Heap                  & 143  & 23  &    100    & \textbf{104} &     84     & 93   & 81   & 78   &\textbf{104}& 49  &     98    &    101    &      101  \\ 
        Loops                 & 727  & 211 &    574    & \textbf{591} &     467    & 380  & 357  & 542  &    575     & 359 &     538   &    544    &      587  \\ 
        ProductLines          & 263  & 19  &\textbf{77}& \textbf{77}  &     56     & 74   & 70   & 74   &\textbf{77} & 48  &     69    &\textbf{77}&\textbf{77}\\ 
        Recursive             & 53   & 25  &    41     & \textbf{45}  &     39     & 21   & 26   & 27   &    43      & 11  &\textbf{45}&    40     &      41   \\ 
        Sequentialized        & 103  & 0   &    79     &      90      &     58     & 35   & 1    & 43   &    75      & 11  &     51    &    57     &\textbf{91}\\ 
        XCSP                  & 119  & 0   &    116    &      107     &\textbf{119}& 102  & 2    & 102  &    118     & 102 &     114   &    96     &      110  \\ 
        Combinations          & 671  & 63  &    238    & \textbf{401} &     167    & 196  & 179  & 224  &    292     & 79  &     338   &    295    &      351  \\ 
        BusyBox & 75   & 0   &    12     &      25      &     6      & 21   & 0    & 0    &    24      & 15  &     19    &    18     &\textbf{29}\\ 
        DeviceDrivers  & 290  & 13  &\textbf{60}&      59      &     6      & 25   & 56   & 47   &    57      & 16  &     42    &    56     &      57   \\ 
        SQLite-MemSafety   & 1    & 0   &    0      &      0       &     0      & 0    & 0    & 0    &    0       & 0   &     0     &    0      &      0    \\ 
        Termination              & 231  & 143 &    212    & \textbf{213} &     195    & 118  & 168  & 145  &    204     & 60  &     179   &    192    &     202   \\ 
        \hline
        {\textit{Cover-Branches}}                          & 3460 & 624 &   1860    & \textbf{2104}&     1406   & 1242 & 1033 & 1487 &    1990    & 896 &     1802  &    1746   &     2075  \\ 
        \hline
    \end{tabular}
    \label{table:CoverBranches}
\end{table*}
\footnotetext{\url{https://test-comp.sosy-lab.org/2022/results/results-verified/}}

{Similarly,} Table~\ref{table:CoverError} {demonstrates demonstrates the error detecting abilities of \textit{FuSeBMC v4}. In particular,} \textit{FuSeBMC} achieved first place in 9 subcategories (\ie \textit{Arrays, BitVectors, ControlFlow, Floats, Heap, Loops, ProductLines, Recursive} and \textit{BusyBox}) reaching the first overall place in this category with the result of {$628$ out of $776$ ($\sim81$\% success rate)}. Overall, the results show that \textit{FuSeBMC} produces \textcolor{black}{test-cases} that detect more security vulnerabilities in C programs than state-of-the-art tools, which successfully {demonstrates that the evaluation objective \textbf{O3} has been achieved}.

\begin{table*}
    \centering
    \caption[Cover-Error]{\textcolor{black}{{\textit{Cover-Error} category results at Test-Comp 2022}{\footnotemark}. The best score for each subcategory is highlighted in bold.}}
    \begin{tabular}{|l||c|c|c|c|c|c|c|c|c|c|c|c|c|}
        \hline
        \multirow{2}{*}{} & & \multicolumn{12}{c|}{Tool} \\
        \cline{3-14}
        \multirow{2}{*}{Subcategory} &
        \rotatebox[origin=c]{90}{Total \# tasks} &
        \rotatebox[origin=c]{90}{CMA-ES Fuzz} &
        \rotatebox[origin=c]{90}{CoVeriTest v2.0.1} &
        \rotatebox[origin=c]{90}{\textit{FuSeBMC} v4.1.14} &
        \rotatebox[origin=c]{90}{ HybridTiger v1.9.2 } &
        \rotatebox[origin=c]{90}{KLEE v2.2} &
        \rotatebox[origin=c]{90}{Legion v1.0} &
        \rotatebox[origin=c]{90}{Legion/SymCC} &
        \rotatebox[origin=c]{90}{LibKluzzer v1.0} & 
        \rotatebox[origin=c]{90}{PRTest v2.2} &
        \rotatebox[origin=c]{90}{Symbiotic v9.0} &
        \rotatebox[origin=c]{90}{Tracer-X v1.2.0} &
        \rotatebox[origin=c]{90}{VeriFuzz v1.2.10} \\
        \hline
        Arrays                & 100 & 0 &  73 & \textbf{99}  & 69  &      89    & 67 & - &      97     & 37  & 74        & 0 & \textbf{99} \\ 
        BitVectors            & 10  & 0 &  8  & \textbf{10}  & 6   &      9     & 0  & - & \textbf{10} & 5   & 8         & 0 & \textbf{10} \\ 
        ControlFlow           & 32  & 0 &  18 & \textbf{32}  & 16  &      27    & 0  & - &      27     & 0   & 24        & 0 &      30     \\ 
        ECA                   & 18  & 0 &  3  &      13      & 1   &      13    & 0  & - &      11     & 0   & 14        & 0 & \textbf{15} \\ 
        Floats                & 33  & 0 &  25 & \textbf{33}  & 23  &      6     & 0  & - &      30     & 3   & 0         & 0 &      32     \\ 
        Heap                  & 56  & 0 &  49 & \textbf{53}  & 42  &      52    & 3  & - & \textbf{53} & 13  &\textbf{53}& 0 & \textbf{53} \\ 
        Loops                 & 157 & 0 &  75 & \textbf{146} & 53  &      95    & 4  & - &      136    & 102 & 81        & 0 &      142    \\ 
        ProductLines          & 169 & 0 & 160 & \textbf{169} & 53  &\textbf{169}& 34 & - & \textbf{169}& 92  & 159       & 0 & \textbf{169}\\ 
        Recursive             & 20  & 0 &  7  & \textbf{19}  & 5   &      16    & 0  & - &      17     & 1   & 17        & 0 &      16     \\ 
        Sequentialized        & 107 & 0 &  61 &      102     & 92  &      86    & 0  & - &      81     & 0   & 79        & 0 & \textbf{104}\\ 
        XCSP                  & 59  & 0 &  50 &      52      & 52  &      37    & 0  & - &      5      & 0   & 41        & 0 & \textbf{55} \\ 
        BusyBox & 13  & 0 &  0  & \textbf{2}   & 0   &      1     & 0  & - &      0      & 0   & 0         & 0 & \textbf{2}  \\ 
        DeviceDrivers  & 2   & 0 &  0  &      0       & 0   &      0     & 0  & - &      0      & 0   & 0         & 0 &      0      \\ \hline 
        {\textit{Cover-Error}}                          & 776 & 0 &  0  & \textbf{628} & 355 &      500   & 57 & - &      528    & 145 & 463       & 0 &      623    \\
        \hline
    \end{tabular}
    \label{table:CoverError}
\end{table*}
\footnotetext{\url{https://test-comp.sosy-lab.org/2022/results/results-verified/}}

\section{Related Work}
\label{sec:RelatedWork}



In this section, we overview related work. Most related techniques fall into one of the following categories: fuzzing, symbolic execution, or a combination of both.

\subsection{Fuzzing}

Barton Miller~\cite{barton1990fault} proposed fuzzing at the University of Wisconsin in the 1990s, and it became the popular software vulnerabilities detection technique~\cite{sutton2007fuzzing}. One of the most common fuzzing tools is American fuzzy lop (AFL)~\cite{bohme2017directed,americanfuzzylop_2021}. AFL is a coverage-based fuzzer that was built to find software vulnerabilities. AFL relies on an evolutionary approach to learn mutations by measuring code coverage. By employing genetic algorithms with guided fuzzing, AFL yields high code coverage. Another tool is LibFuzzer~\cite{serebryany2015libfuzzer}, which uses code coverage information generated by LLVM's (SanitizerCoverage) instrumentation to produce \textcolor{black}{test-cases}. LibFuzzer is best suited for testing libraries that have small input with a run-time of milliseconds for each input to guarantee not crashing on invalid input in library code\footnote{\url{https://llvm.org/docs/LibFuzzer.html\#q-so-what-exactly-this-fuzzer-is-good-for}}. Vuzzer~\cite{rawat2017vuzzer} is a fuzzer with an application-aware strategy. The main advantage of this strategy is that it does not need any knowledge of the application or input format in advance.

To maximize coverage and explore deeper paths, the tool leverages control- and data-flow features based on static and dynamic analysis to infer fundamental properties of the application. This enables a much faster generation of interesting inputs than an application-agnostic approach. Wang et al.~\cite{wang2017skyfire} proposed an approach that utilizes data-driven seed generation. It relies on extracting the knowledge of grammar to process and generate well-distributed seed inputs for fuzzing programs. Skyfire is designed for probabilistic context-sensitive grammar (PCSG) to identify syntax features and semantic rules. AFLFast~\cite{bohme2017coverage} is an enhanced version of AFL applying various strategies to exercise a low-frequency path. The tool achieved a $7$x speedup over AFL~\cite{bohme2017coverage}. 

GTFuzz~\cite{li2020gtfuzz} is a tool that prioritizes inputs based on extracting syntax tokens that guard the target place 
The backward static analysis technique 
extracts these tokens. Also, GTFuzz benefits from this extraction by improving the mutation algorithm. Smart grey-box fuzzing (SGF)~\cite{pham2019smart} is a fuzzer that employs high-level structural representation on the original seeds to generate high-impact seeds. Similarly, AFLSmart~\cite{pham2019smart} is a structure-aware fuzzer that combines the PEACH fuzzing engine with the AFL fuzzer. Instrim~\cite{hsu2018instrim} is a Control Flow Graph (CFG) -aware fuzzer. It analyses the CFG of the PUT in an attempt to instrument fewer code blocks, thereby speeding up the fuzzing. AutoFuzz~\cite{gorbunov2010autofuzz} is a tool that utilizes fuzzing to verify network protocols. It begins with finding the protocol specifications and then finding vulnerabilities by using fuzzing 
{Also, there is Peach Fuzzer~\cite{zhang2015peach} is an approach that sends random input to a PUT in an attempt to find security vulnerabilities. It is frequently used to detect security vulnerabilities in input validation and application logic.} Various other fuzzers and fuzzing techniques have been developed, each with unique features. For example, directed greybox fuzzing ~\cite{bohme2017directed} uses simulated annealing in an attempt to guide the fuzzer to explore a particular section of the program under test. SYMFUZZ~\cite{cha2015program} controls the selection of paths, while Alexandre Rebert's approach~\cite{rebert2014optimizing} uses guided seed selection.

One of the weaknesses of pure fuzzing approaches is their inability to find \textcolor{black}{test-cases} that explore program code beyond complex guards, as they essentially work by randomly mutating seeds and, therefore, struggle to find inputs that satisfy the guards.

\subsection{Symbolic Execution {and Bounded Model Checking}}

Symbolic execution and BMC have shown competence in producing high-coverage \textcolor{black}{test-cases} and detecting errors in complex software. One of the more popular symbolic execution engines is KLEE~\cite{cadar2008klee}. KLEE is a tool that explores the search space path-by-path by utilizing LLVM compiler infrastructure and dynamic symbolic execution. KLEE has been utilized in many specialized tools for its reliability as a symbolic execution engine. 
{For example, Symbiotic 8~\cite{10.1007/978-3-030-72013-1_31} symbolic execution tool is built on top of KLEE while adding plugins, such as Predator~\cite{dudka2011predator} and DG~\cite{chalupa2020dg}, to fulfill different functions. The latest addition to the tool is Slowbeast~\cite{Slowbeast}, which incorporates the k-induction technique into symbolic execution.}
Furthermore, {Tracer~\cite{jaffar2012tracer} is a verification tool that uses constraint logic programming (CLP) and interpolation methods.} Another tool based on symbolic execution is DART~\cite{godefroid2005dart}. It conducts software analysis and applies automatic random testing to find software bugs. BAP~\cite{brumley2011bap} is developed on top of Vine~\cite{song2008bitblaze}, which relies on symbolic execution. BAP has useful analysis and verification techniques. BAP relies on an intermediate language (IL) in its analysis. {Also, SymbexNet~\cite{song2014symbexnet} and SymNet~\cite{sasnauskas2012integration} are for verification of network protocols implementation.} Avgerinos, Thanassis, et al.~\cite{avgerinos2014enhancing} presented an approach that enhances symbolic execution with verification-based algorithms. It works to increase the effect of dynamic symbolic execution. The approach showed its ability to detect bugs and achieve higher code coverage than other dynamic symbolic execution approaches. CoVeriTest~\cite{beyer2019coveritest} is a Cooperative Verifier Test generation that utilizes a hybrid approach for test generation. It applies different conditional model checkers in iterations with many value-analysis configurations. CoVeriTest also changes the level of cooperation and assigns the time budget of each verifier. However, symbolic execution suffers from the path explosion problem related to loops and arrays, impacting its practicality.

\subsection{Combination}

The combination of symbolic execution and BMC with fuzzing has been used recently to combine the strengths of both techniques. For example, VeriFuzz~\cite{chowdhury2019verifuzz} is a state-of-the-art tool we have previously compared to \textit{FuSeBMC}. The authors describe it as a program-aware fuzz tester that combines feedback-driven evolutionary fuzz testing with static analysis. It also employs grey-box fuzzing to exploit lightweight instrumentation for observing the behaviors that occur during test runs. VeriFuzz earned first place in Test-Comp 2020~\cite{beyer2020second}. Driller~\cite{stephens2016driller} is a hybrid vulnerability excavation tool developed by Stephens et al. It finds deeply embedded bugs by leveraging guided fuzzing and concolic execution in a complementary manner. It employs concolic execution to analyze the program and trace the inputs. Concolic execution also guides fuzzing on different paths by using a constraint-solving engine.
Stephens et al. combine the strengths of the two techniques and mitigate the weaknesses to avoid path explosion in concolic analysis. First, Driller divides the application, based on checks of particular values of a specific input, into \emph{compartments}. Then by utilizing the proficiency of fuzzing, Driller explores possible values of general input in a compartment. Although Driller showed its efficiency in detecting more vulnerabilities, it may drive to path explosion problems because it requires a lot of computing power. MaxAFL~\cite{kim2020maxafl} is best described as a gradient-based fuzzer built on top of AFL. First, the developer finds the Maximum Expectation of Instruction Count (MEIC) using lightweight static analysis. Then, using MEIC, they generate an objective function and apply a gradient-based optimization algorithm to generate efficient inputs by minimizing the objective function. Hybrid Fuzz Testing~\cite{pak2012hybrid} is a tool that generates provably random \textcolor{black}{test-cases} efficiently such that it guarantees the execution of unique paths. {Moreover, it finds unique execution paths} by using symbolic execution to find frontier nodes that lead such paths. The tool also collects all possible frontier nodes depending on resource constraints to employ fuzzing with provably random input, preconditioned to lead to each frontier node.

He et al.~\cite{he2019learning} proposed an approach to learning a fuzzer from symbolic execution. First, it phrases the learning task in the framework of imitation learning. {Then, it employs} symbolic execution to generate quality inputs with high coverage while a fuzzer learns using neural networks to be used to fuzz new programs. Badger\cite{noller2018badger} provides a hybrid testing approach for complexity analysis. It generates new input using Symbolic PathFinder~\cite{puasuareanu2010symbolic} and provides the Kelinci fuzzer with worst-case analysis. Badger aims to increase coverage and resource-related cost for each path by using fuzz testing. LibKluzzer~\cite{le2020llvm} is a novel implementation combining Symbolic execution and fuzzing. Its strength resides in the fusion of coverage-guided fuzzing and white-box fuzzing strengths. LibKluzzer is constructed of LibFuzzer, and an extension of KLEE called KLUZZER~\cite{le2019kluzzer}. Munch~\cite{ognawala2018improving}is an open-source framework hybrid tool. It employs fuzzing with seed inputs generated by symbolic execution and targets symbolic execution when fuzzing saturates. It aims to reduce the number of queries to the SMT solver to focus on the paths that may reach uncovered functions. The developers created Munch to improve function coverage. 
FairFuzz~\cite{lemieux2018fairfuzz} is a grey-box fuzzer that utilizes guided-mutation. It uses coverage to achieve the guidance by employing a mutation mask for every pair of seeds and the rare branches to direct the fuzzing to reach each rare branch. SAFL~\cite{wang2018safl} is an efficient fuzzer for C/C++ programs. It utilizes symbolic execution (in a lightweight approach) to generate initial seeds that can get an appropriate fuzzing direction. \textcolor{black}{{SAGE (Scalable Automated Guided Execution) proposed by Godefroid et al.~\cite{godefroid2012sage} is a hybrid fuzzer developed at Microsoft Research. Microsoft uses SAGE extensively, where it has successfully found many security-related bugs. It employs generational search to extend dynamic symbolic execution and increase code coverage by negating and solving the path predicates. Also, SAGE relies on the random test style used by DART to mutate good inputs using grammar. }}

Overall, the combination of Fuzzers and Symbolic execution to verify software has been successful. Our approach builds on this combination by utilizing smart seed generation, the Tracer subsystem, and other unique features as described above.

\section{Conclusion}

In this paper, we presented \textit{FuSeBMC} v4, a test generator that relies on smart seed generation to improve the state-of-the-art in hybrid fuzzing and achieve high coverage for C programs. First, \textit{FuSeBMC} analyses and injects goal labels into the given C program. Then, it ranks these goal labels according to the given strategy. After that, the engines are employed to produce smart seeds for a short time. Then, \textit{FuSeBMC} \ahmedhappy{starts test generation by running a bounded model checking engine along with fuzzers}. This Tracer will generally manage the tool to record the goals covered and deal with the transfer of information between the engines by providing a shared memory to harness the power and take advantage of the power of each engine. 
So, the BMC engine helps give the seed that prevents the fuzzing engine from struggling with complex mathematical guards. Furthermore, Tracer dynamically evaluates \textcolor{black}{test-cases} to convert high-impact cases into seeds for subsequent test fuzzing. This approach was evaluated by participating in the fourth international competition on software testing Test-Comp 2022. Our approach \textit{FuSeBMC} showed its effectiveness by achieving first place in the \textit{Cover-branches} category, first place in the \textit{Cover-Error} category, and first place in the \textit{Overall} category. This performance is due to various features of our tool, the most important of which are the following. First, the generation of smart seeds, which help harness the power of the fuzzers and allow them to fuzz deeper. Second, simplifying the target program by limiting the bounds of potentially infinite loops to avoid the path explosion problem and produce seeds faster. Third, utilizing the static analysis to manage the mutation process by limiting the range of values input variables can take, speeding up the fuzzing process. In the future, we plan on developing a tool to deal with different types of programs, such as multi-threaded programs. Furthermore, we work with the SCorCH project\footnote{\url{https://scorch-project.github.io/about/}} to improve our performance in detecting memory safety bugs by incorporating SoftBoundCETS ~\cite{nagarakatte2009softbound} into \textit{FuSeBMC}.

\newpage
\label{lastpage}

\bibliographystyle{ACM-Reference-Format}

\end{document}